\documentclass[aps,prl,twocolumn,superscriptaddress]{revtex4}
\usepackage[dvipdfm]{graphicx}

\begin{document}






\title{Soft X-ray Absorption and Photoemission Studies of 
Ferromagnetic Mn-Implanted 3$C$-SiC}

\author{Gyong Sok Song}
\affiliation{Department of Physics and Department of Complexity Science and Engineering, University of Tokyo, 7-3-1, Hongo, Tokyo, 113-0033}
\author{Takashi Kataoka}
\affiliation{Department of Physics and Department of Complexity Science and Engineering, University of Tokyo, 7-3-1, Hongo, Tokyo, 113-0033}
\author{Masaki Kobayashi}
\affiliation{Department of Physics and Department of Complexity Science and Engineering, University of Tokyo, 7-3-1, Hongo, Tokyo, 113-0033}
\author{Jong Il Hwang}
\affiliation{Department of Physics and Department of Complexity Science and Engineering, University of Tokyo, 7-3-1, Hongo, Tokyo, 113-0033}
\author{Masaru Takizawa}
\affiliation{Department of Physics and Department of Complexity Science and Engineering, University of Tokyo, 7-3-1, Hongo, Tokyo, 113-0033}
\author{Atsushi Fujimori}
\affiliation{Department of Physics and Department of Complexity Science and Engineering, University of Tokyo, 7-3-1, Hongo, Tokyo, 113-0033}
\author{Takuo Ohkochi}
\affiliation{Synchrotron Radiation Research Unit, Japan Atomic Energy Agency, Sayo-gun, Hyogo 679-5148}
\author{Yukiharu Takeda}
\affiliation{Synchrotron Radiation Research Unit, Japan Atomic Energy Agency, Sayo-gun, Hyogo 679-5148}
\author{Tetsuo Okane}
\affiliation{Synchrotron Radiation Research Unit, Japan Atomic Energy Agency, Sayo-gun, Hyogo 679-5148}
\author{Yuji Saitoh}
\affiliation{Synchrotron Radiation Research Unit, Japan Atomic Energy Agency, Sayo-gun, Hyogo 679-5148}
\author{Hiroshi Yamagami}
\affiliation{Synchrotron Radiation Research Unit, Japan Atomic Energy Agency, Sayo-gun, Hyogo 679-5148}
\affiliation{Department of Physics, Faculty of Science, Kyoto Sangyo University,Kyoto 603-8555}
\author{Fumiyoshi Takano}
\affiliation{Nanotechnology Research Institute, National Institute of Advanced Industrial Science and Technology, 1-1-1 Umezono, Tsukuba, Ibaraki 305-8568}
\author{Hiro Akinaga}
\affiliation{Nanotechnology Research Institute, National Institute of Advanced Industrial Science and Technology, 1-1-1 Umezono, Tsukuba, Ibaraki 305-8568}

\begin{abstract}

We have performed x-ray photoemission spectroscopy (XPS), x-ray absorption spectroscopy (XAS), and resonant photoemission spectroscopy (RPES) measurements of Mn-implanted 3$C$-SiC (3$C$-SiC:Mn) and carbon-incorporated Mn$_{5}$Si$_{2}$ (Mn$_{5}$Si$_{2}$:C). The Mn 2$p$ core-level XPS and XAS spectra of 3$C$-SiC:Mn and Mn$_{5}$Si$_{2}$:C were similar to each other and showed gintermediateh behaviors between the localized and itinerant Mn 3$d$ states. 
 The intensity at the Fermi level was found to be suppressed in 3$C$-SiC:Mn compared with Mn$_{5}$Si$_{2}$:C. These observations are consistent with the formation of Mn$_{5}$Si$_{2}$:C clusters in the 3$C$-SiC host, as observed in a recent transmission electron microscopy study. 
\end{abstract}

\maketitle

KEYWORDS: diluted magnetic semiconductor, SiC, silicide, electronic structure, photoemission spectroscopy, x-ray absorption spectroscopy
\vspace{.4cm}

In recent years, there has been growing interest in the possibility of using electron spins in electronic devices for the transfer, processing, and storage of information. 
In such prospects for the realization of spintronic devices, diluted magnetic semiconductors (DMSs) are key materials and remarkable development has been achieved after the discovery of the relatively high Curie temperature ($T\mathrm{_C}\sim$110 K) in Ga$_{1-x}$Mn$_x$As.\cite{Matsukura} 
Inspired by the theoretical predictions toward DMSs with higher $T\mathrm{_C}$'s,\cite{Dietl1, Ksato_ZnO1, Ksato_GaN1} various experimental studies on wide-gap DMSs, especially ZnO and GaN, have been performed and many wide-gap DMSs have been reported to show ferromagnetism at room temperature.\cite{ZnCoO_Tabata, GaN_Munekata, Sonoda, HashimotoJCG03} However, only few experimental studies on silicon carbide (SiC)-based DMS have been reported \cite{Theodoropoulou, Syvajarvi} in spite of the high potential of SiC for high-power devices because of its wide band gap, high thermal conductivity, high hardness, and chemical inertness. 

Recently, Takano $et$ $al$.\cite{Takano_3C} have reported ferromagnetism in Mn-implanted 3$C$-SiC (3$C$-SiC:Mn) and a $T\mathrm{_C}$ of 245 K has been observed. 
In their report, high resolution transmission electron microscopy (HRTEM) and selected-area diffraction (SAD) measurements have revealed that cluster precipitations occurred in 3$C$-SiC:Mn and that electron diffraction from the cluster region showed superposed patterns of Mn$_{5}$Si$_{2}$ and 3$C$-SiC. The Mn content was estimated to be 20 \% in the cluster region and below 1 \% in the 3$C$-SiC host region from energy dispersive x-ray measurements. Since Mn$_{5}$Si$_{2}$ is known as a paramagnetic metal,\cite{Mn5Si2_para} the ferromagnetism of the 3$C$-SiC:Mn sample cannot be explained simply by the existence of Mn$_{5}$Si$_{2}$. On the other hand, a Mn$_{5}$Si$_{2}$-derived ferromagnetic material generated by thermal reaction between Mn and 4$H$-SiC has been recently reported,\cite{Takano_4H} and the $T\mathrm{_C}$ of 300 K and the ferromagnetic hysteresis have been observed. HRTEM and SAD analysis were also performed on this material and the diffraction from the reacted region showed a clear pattern of Mn$_{5}$Si$_{2}$. Because the ferromagnetic hysteresis disappeared when the Mn$_{5}$Si$_{2}$-derived region was removed by chemical etching, it was proposed that carbon incorporation into the paramagnetic Mn$_{5}$Si$_{2}$ caused the ferromagnetism.\cite{Wang_4H} The incorporation of carbon atoms was confirmed by secondary ion mass spectrometry and also by C 1$s$ core-level photoemission. 
It has remained unclear, however, whether the ferromagnetism of 3$C$-SiC:Mn comes from dilute Mn atoms in the SiC host (SiC:Mn DMS) or the clusters of carbon-incorporated Mn$_{5}$Si$_{2}$ (Mn$_{5}$Si$_{2}$:C). The characterization of the electronic structure using high energy spectroscopic methods such as photoemission spectroscopy and x-ray absorption spectroscopy will bring us more direct information.

In this study, we present the results of x-ray photoemission spectroscopy (XPS), x-ray absorption spectroscopy (XAS), and resonant photoemission spectroscopy (RPES) measurement of 3$C$-SiC:Mn. 
For comparison, those of Mn$_{5}$Si$_{2}$:C are also presented.
The electronic structure of Mn and a possible origin of the ferromagnetism in 3$C$-SiC:Mn are discussed. 

3$C$-SiC:Mn was fabricated by Mn-implantation into a 3$C$-SiC (001) wafer with a 5 $\mu$m homoepitaxial layer.
The epitaxial layer showed $n$-type conduction with the carrier density of 1$\times$10$^{16}$ cm$^{-3}$. The Mn implantation energy of 350 keV was employed and the dose was 1$\times$10$^{16}$ cm$^{-2}$. During the implantation, the wafer was maintained at 800 $^{\circ}$C to prevent amorphization. The implanted wafer was annealed at 1650 $^{\circ}$C for 20 min in an Ar atmosphere. Details of the sample preparation and characterization were described elsewhere.\cite{Takano_3C}
Mn$_{5}$Si$_{2}$:C was fabricated by depositing Mn metal on a 4$H$-SiC homoepitaxial wafer followed by annealing. As the first step, the SiC wafer was thermally cleaned at 1000 $^{\circ}$C for 10 min in a high vacuum chamber ($\sim1.3\times10^{-6}$ Pa during cleaning) to remove thin oxide layer at the surface. After the cleaning, Mn with a thickness of $\sim$50 nm was deposited on the surface using a Knudsen cell in an ultrahigh vacuum chamber at a substrate temperature of 300 $^{\circ}$C. As the last step, annealing was performed at 1000 $^{\circ}$C for 3 min. Details of the sample preparation and characterization were described in elsewhere.\cite{Takano_4H}

XPS measurements were performed using a Gammadata Scienta SES-100 hemispherical analyzer and a Mg-$K\alpha$ source ($h\nu$ = 1253.6 eV). The XAS and RPES measurements were performed at BL23SU of SPring-8 using photons of $h\nu$ = 630 - 660 eV. Photoelectrons were collected by a Gammadata Scienta SES-2002 hemispherical analyzer. All the spectra were taken at room temperature. The total resolution of the spectrometer including thermal broadening was $\sim$800 meV for XPS and $\sim$200 meV for RPES, respectively. The monochromator resolution was $E$/$\Delta$$E$ $>$10,000. The absorption spectra were measured by the total electron yield method. The base pressure was below 3$\times$10$^{-8}$ Pa. Prior to the measurements, Ar$^{+}$-ion sputter etching was performed for surface cleaning. The Ar pressure was 1$\times$10$^{-5}$ Pa and the incident angle of the Ar$^{+}$-ion beam was fixed at and 45$^{\circ}$ from the surface normal. The acceleration voltage was chosen as 1.5 kV for XPS and 3 kV for XAS and RPES.

Figure 1 shows the Mn 2$p$ core-level XPS spectra of 3$C$-SiC:Mn and Mn$_{5}$Si$_{2}$:C. For comparison, the spectrum of Mn metal and that of an oxidized surface of Mn$_{5}$Si$_{2}$:C which was taken before sputtering are also shown. Here, the spectra have been normalized to the maximum intensity of the Mn 2$p_{3/2}$ peak. The two-peak structure due to the Mn 2$p_{3/2}$ and 2$p_{1/2}$ spin-orbit splitting is seen in every spectrum. 
One can see that the spectrum of 3$C$-SiC:Mn is similar to that of Mn$_{5}$Si$_{2}$:C and also that of Mn metal. The peak positions of the Mn 2$p_{3/2}$ peaks of 3$C$-SiC:Mn, Mn$_{5}$Si$_{2}$:C, and Mn metal are almost the same at 639 eV, but differ from that (642 eV) of the oxidized surface. Not only the energy positions but also the line shapes are similar between 3$C$-SiC:Mn, Mn$_{5}$Si$_{2}$:C, and Mn metal and obviously differ from those of the oxidized surface. The asymmetric tails appearing on the higher binding energy ($E_B$) side of the peaks are well known as a characteristic feature of metallic systems, caused by the screening of the core hole by conduction electrons.\cite{Hufner} On the contrary, the peak of the oxidized surface does not show an asymmetric line shape and is relatively broad. If the Mn ions doped in 3$C$-SiC substitute for the Si and/or C sites, namely, the Mn atoms are tetrahedrally coordinated by C and/or Si atoms, the spectrum would show a spectral line shape similar to those of Zn$_{1-x}$Mn$_{x}$O,\cite{Mizokawa_ZnMnO} Ga$_{1-x}$Mn$_{x}$As,\cite{Okabayashi_PES} and Ga$_{1-x}$Mn$_{x}$N.\cite{Hwang_XPS} This indicates that both Mn atoms in 3$C$-SiC:Mn and in Mn$_{5}$Si$_{2}$:C are not tetrahedrally coordinated by C and/or Si atoms and have local electronic structures which are relatively similar to that of Mn metal.
We note that there is a slight broadening of the asymmetric tails of the Mn 2$p$ peaks for 3$C$-SiC:Mn and Mn$_{5}$Si$_{2}$:C compared to that for Mn metal. Similar broadening was observed in the case of the Mn:Ge system.\cite{MnGe_Verdini}  In this reference, hybridization between the Mn 3$d$ orbitals and Ge valence orbitals, namely, covalency was proposed as the origin of the broadening of the Mn 2$p$ core-level photoemission spectrum compared to that of Mn metal. Therefore, it is likely that Mn - Si or Mn - C hybridization causes the slight peak broadening observed in the present case.


\begin{figure}[tbp]
\begin{center}
\includegraphics[width=5.5cm]{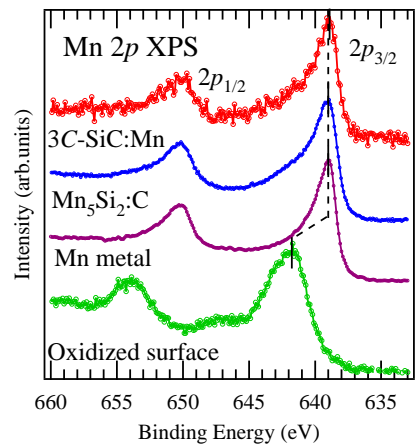}
\caption{(Color online) Mn 2$p$ core-level spectra of 3$C$-SiC:Mn and Mn$_{5}$Si$_{2}$:C. Dashed lines indicates the position of the 2$p_{3/2}$ peak. For comparison, the spectra of Mn metal and Mn$_{5}$Si$_{2}$:C taken before sputtering (oxidized surface) are shown.}
\label{XPS_SCM}
\end{center}
\end{figure}

Figure 2 shows the Mn 2$p$ XAS spectra of 3$C$-SiC:Mn and Mn$_{5}$Si$_{2}$:C, together with that of Mn metal and that of oxidized surface of Mn$_{5}$Si$_{2}$:C taken before sputtering. Here, the spectra have been aligned at the main peak, and the intensities of these spectra have been normalized to the maximum intensity of the main peak. The fine structures observed for the spectrum of the oxidized surface show multiplet structures characteristic of the localized 3$d$ states of the Mn$^{2+}$ ion.\cite{VanderLaan} As in the case of the XPS result, the XAS spectra of 3$C$-SiC:Mn and Mn$_{5}$Si$_{2}$:C have almost the same line shapes and are similar to that of Mn metal, indicating that the electronic structures of Mn in 3$C$-SiC:Mn and Mn$_{5}$Si$_{2}$:C are rather similar to that of Mn metal. 
The spectra, however, show weak peaks and shoulders indicated by dashed vertical lines. These structures may represent weakly localized states of the Mn$^{2+}$ ion. In the system of Mn:Ge,\cite{MnGe_Verdini} too, in addition to the globally metal-like line shape of Mn 2$p$ XAS, weak multiplet structures were observed and were considered as a result of hybridization between Mn and Ge. Carbone $et$ $al$.\cite{Carbone_MnSi} also suggested that covalency in MnSi, which is a helimagnetic metal, could be the origin of the multiplet structures superposed on the metallic feature of the Mn 2$p$ XAS spectrum. Therefore, it is likely that Mn 3$d$ electrons in 3$C$-SiC:Mn and Mn$_{5}$Si$_{2}$:C are gintermediateh between the localized and itinerant limits. We note that another possible reason for the observed intermediate behavior is the coexistence of metallic Mn and residual surface Mn oxides. In fact, a small amount of oxygen atoms were found in the O 1$s$ core-level spectra (13\% and 19\% of the total number of atoms for 3$C$-SiC:Mn and Mn$_{5}$Si$_{2}$:C, estimated from the intensities of the Mn 2$p$, Si 2$p$, C 1$s$ and O 1$s$ core-level spectra). 
However, since the Mn 2$p$ XPS spectra of 3$C$-SiC:Mn and Mn$_{5}$Si$_{2}$:C were quite different from that of the oxidized surface, we believe that most of the Mn atoms in 3$C$-SiC:Mn and Mn$_{5}$Si$_{2}$:C were not oxidized. 
Thus, one can conclude that the electronic structure of Mn in 3$C$-SiC is similar to that in Mn$_{5}$Si$_{2}$:C, consistent with the result of the TEM study which has demonstrated the existence of Mn$_{5}$Si$_{2}$-derived clusters in 3$C$-SiC.\cite{Takano_3C}

\begin{figure}[tbp]
\begin{center}
\includegraphics[width=5.5cm]{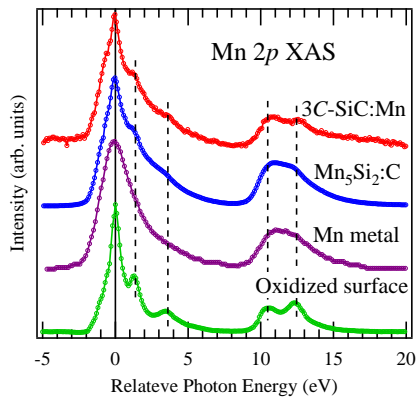}
\caption{(Color online) Mn 2$p$ XAS spectra of 3$C$-SiC:Mn and Mn$_{5}$Si$_{2}$:C. For comparison, the spectra of Mn metal and Mn$_{5}$Si$_{2}$:C taken before sputtering (oxidized surface) are shown. Dashed vertical lines indicate the peak or shoulder positions.}
\label{XAS_SCM}
\end{center}
\end{figure}

We have also studied the valence-band electronic states by Mn 2$p$-3$d$ RPES measurements. RPES is a powerful tool to investigate the electronic structure of doped Mn ions in DMS because the resonance enhancement of the Mn 3$d$ partial density of states (PDOS) enables us to extract it from the valence band.\cite{Mizokawa_ZnMnO, Hwang_XPS, Okabayashi_RPES} Figure 3 shows the Mn 2$p$-3$d$ RPES spectra of 3$C$-SiC:Mn and Mn$_{5}$Si$_{2}$:C. 
The RPES spectra of 3$C$-SiC:Mn shown in panel\,(a) exhibit only weak resonant enhancement because of the low Mn content in 3$C$-SiC (below 1 at.\% Mn estimated from the intensities of the Si 2$p$, C 1$s$, and Mn 2$p$ core-level XPS spectra). 
Nevertheless, a difference spectrum has been clearly obtained by subtracting the off-resonance spectrum A from the on-resonance spectrum B as shown at the bottom of panel\,(a). 
The RPES spectra of Mn$_{5}$Si$_{2}$:C shown in panel\,(b) clearly exhibit much stronger resonance enhancement as well as the Fermi edge, but are dominated by the Mn $L_{3}VV$ Auger peaks, move toward higher $E_B$ with increasing photon energy because Auger electrons have constant kinetic energies. The strong Auger signal is a typical feature of a metallic material.
The appreciable resonance enhancement of the Fermi edge in panel\,(b) indicates that the conduction electrons have sizable amount of Mn 3$d$ character, whereas 
\begin{figure}[bp]
\begin{center}
\includegraphics[width=9cm]{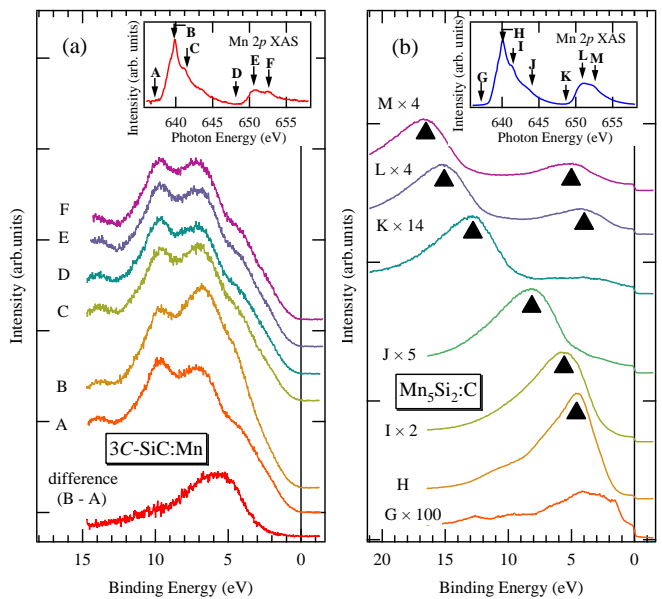}
\caption{(Color online) Comparison of the valence-band photoemission spectra of 3$C$-SiC:Mn (a) and Mn$_{5}$Si$_{2}$:C (b) taken at various photon energies in the Mn 2$p$-3$d$ core-excitation region. Black triangles represent the Mn $L_{3}VV$ Auger signal. Insets in both panels show the corresponding Mn 2$p$ XAS spectra. Arrows indicate the photon energy where the valence-band spectra were taken.}
\label{RPES2_SCM}
\end{center}
\end{figure}

we could not observe detectable intensity at $E\mathrm{_F}$ of 3$C$-SiC:Mn as shown in panel\,(a). 
In general, the electronic structure of a metal cluster including the intensity at $E\mathrm{_F}$ is size-dependent. It is considered that finite size effects become observable for metal particles typically smaller than 10 nm.\cite{Kubogap} For example, theory of quantum size effect (Kubo effect) predicts the opening of a gap with decreasing cluster size.\cite{Kubogap} 
It has also been reported that final-state effects of photoemission process leads to the disappearance of the Fermi edges in small metal clusters.\cite{Hovel_cluster}
Therefore, the apparently different RPES spectra of 3$C$-SiC:Mn and Mn$_{5}$Si$_{2}$:C are compatible with the similar Mn 2$p$ XPS and XAS spectra.

The present results suggest that the local electronic structure of Mn in 3$C$-SiC is similar to that of Mn in Mn$_{5}$Si$_{2}$:C and different from the localized states which is expected for a transition-metal atom at the substitutional site of SiC. 
 Furthermore, a blocking phenomenon and the blocking temperature of 215 K were observed from the field-cooling and zero-field-cooling magnetization measurements of the present 3$C$-SiC:Mn sample (not shown), suggesting that the ferromagnetic-like behavior of 3$C$-SiC:Mn comes from a group of clustered ferromagnets. 
Thus, it is considered that Mn$_{5}$Si$_{2}$:C clusters are responsible for the ferromagnetic properties of 3$C$-SiC:Mn. 
Since most of the implanted Mn atoms formed silicide clusters in the present study, the content of substitutional Mn in SiC may be small. A recent density functional calculation\cite{Miao_PRB06} has pointed out that substitutional Mn atoms at the Si site prefer long-range ferromagnetic coupling in 3$C$- and 4$H$-SiC. Fabrication of SiC by molecular-beam epitaxy with high crystallinity have been reported\cite{SiC_MBE} and an incorporation of a substantial amount of Mn into substitutional sites using other techniques such as a low-temperature-molecular-beam epitaxy may be an interesting possible route to fabricate SiC-based ferromagnetic semiconductors. 

We have performed Mn 2$p$ XPS, XAS, and Mn 2$p$-3$d$ RPES studies of 3$C$-SiC:Mn and Mn$_{5}$Si$_{2}$:C. The Mn 2$p$ XPS and XAS of 3$C$-SiC:Mn show gintermediateh feature, similar to Mn$_{5}$Si$_{2}$:C. 
In Mn 2$p$-3$d$ RPES, however, a clear Fermi edge of Mn 3$d$ character has been observed for Mn$_{5}$Si$_{2}$:C while no Fermi edge was observed for 3$C$-SiC:Mn. Cluster size effects may explain the suppression of the intensity at $E\mathrm{_F}$ in 3$C$-SiC:Mn. 
These observations are consistent with the existence of Mn$_{5}$Si$_{2}$-derived clusters in 3$C$-SiC reported in the TEM study. It is possible that the Mn$_{5}$Si$_{2}$:C clusters are responsible for the ferromagnetism of 3$C$-SiC:Mn.

\vspace{.4cm}
This work was supported by a Grant-in-Aid for Scientific Research in Priority Area gCreation and Control of Spin Currenth(19048012) from MEXT, Japan. The experiment at Spring-8 was approved by the Japan Synchrotron Radiation Research Institute (JASRI) Proposal Review Committee (Proposal No. 2007A3832).

\end{document}